\documentclass[]{aa} 
\usepackage{graphicx}
\begin{document}
   \title{Age distribution of the central stars of galactic disk planetary nebulae}
   \titlerunning{Age distribution of CSPN}

%   \subtitle{I. Overviewing the $\kappa$-mechanism}

   \author{W. J. Maciel
          \inst{}
          R. D. D. Costa
          \inst{}
          \and
          T. E. P. Idiart
          \inst{}
          }

   \institute{Astronomy Department, IAG/USP, University of S\~ao Paulo, Rua do Mat\~ao 1226,
              05508-900, S\~ao Paulo SP, Brazil\\
              \email{maciel@astro.iag.usp.br, roberto@astro.iag.usp.br, thais@astro.iag.usp.br}}

   \date{Received ...; accepted ...}

% \abstract{}{}{}{}{} 
% 5 {} token are mandatory
 
  \abstract
   {The determination of ages of central stars of planetary nebulae (CSPN) is a 
    complex problem, and there is presently no single method that can be generally applied. 
    We  have developed several methods to estimate the ages of CSPN, based both on 
    the observed nebular properties and in some properties of the stars themselves.}
   {Our aim is to estimate the ages and the age distribution of  CSPN and to compare
   the derived results with mass and age determinations of CSPN and white dwarfs
   based on empirical determinations of these quantities.}
   {We discuss several methods to derive the age distribution of CSPN, namely,
   (i) the use of an age-metallicity relation that also depends on the galactocentric 
   distance, (ii) the use of an age-metallicity relation obtained for the galactic disk,
   and  (iii) the determination of ages from the central star masses obtained from the 
   observed nitrogen abundances.}
    {We consider a sample of planetary nebulae in the galactic disk, most of 
    which ($\sim$ 69\%) are located in the solar neighbourhood, within 3 kpc from
    the Sun. We estimate the age distribution of CSPN with average uncertainties
    of 1-2 Gyr, and compare our results with the expected distribution based both 
    on the observed mass distribution of white dwarfs and on the age distribution 
    derived from available mass distributions of CSPN. }
   {We conclude most CSPN in the galactic disk have ages under 6 Gyr, and that the age 
    distribution is peaked around 2-4 Gyr.}


   \maketitle
%
%-----------------------------------------------------------------------------------------------------
\section{Introduction}  % section 1
%-----------------------------------------------------------------------------------------------------

Planetary nebulae (PN) are the offspring of intermediate mass stars with main sequence masses 
between 0.8 and 8~$M_\odot$, approximately. As a consequence, their properties reflect different 
physical conditions depending on the masses -- and therefore ages -- of their  central stars (CSPN), 
which makes these objects extremely important in the study of the chemical evolution of the Galaxy  
(see for example Maciel et al. \cite{mlc2006}). As an example, some recent theoretical models predict 
a time flattening of the observed radial abundance gradients in the galactic disk, while other models 
predict just the opposite behaviour. This can be analysed on the basis of abundance determinations in 
PN or in open cluster stars. In both cases, the results depend critically on the adopted ages of the 
objects considered.

The determination of ages of CSPN is a complex problem, and there is presently no single method that 
can be generally applied. In fact, most accurately determined ages refer to relatively young objects,
for which methods such as lithium depletion, activity or cluster membership can be applied (see 
for example Hillebrand et al. \cite{hillebrand}). Our group has pioneered in the treatment of this 
problem, and we have developed several methods to estimate the ages of the PN progenitor stars, based 
both on the observed nebular properties and in some properties of the stars themselves (cf. Maciel et 
al. \cite{mcu2003}, \cite{mlc2005a}, \cite{mlc2006}, \cite{mkc2008}). According to Soderblom 
(\cite{soderblom}), most age determination methods can be classified as (i) fundamental, (ii) model
dependent, (iii) empirical, and (iv) statistical. The methods discussed in this paper belong to the
last class. In principle, the  traditional methods to derive the ages of galactic stars can be applied 
to CSPN, such as the use of theoretical isochrones (see for example Idiart et al. \cite{imc2007}), 
particularly for extragalactic nebulae. On the other hand, the physical properties of these objects 
are not as well known as in the case of normal stars, so that the derived isochrones are generally 
uncertain, leading to the need of alternative methods. 

In this work, we discuss three methods developed so far by our group, namely the use of 
(i) an age-metallicity relation that depends on the galactocentric distance  (Section 2.1a),
(ii) a simpler age-metallicity relation for the galactic disk (Section 2.1b), and (iii) the 
determination of ages from the central star masses obtained directly from the observed nitrogen 
abundance (Section 2.2). In Section~3, we estimate the expected age distribution of the CSPN 
based on (i) the observed mass distribution of white dwarf stars, and (ii) available mass 
distributions of CSPN, and compare the results with the distributions obtained by the methods 
mentioned above. In Section~4 a discussion is given, followed by some conclusions.

%-----------------------------------------------------------------------------------------------------
\section{Age determination and distribution of CSPN progenitors from nebular abundances}  % section 2
%-----------------------------------------------------------------------------------------------------

%.........................................................................................Section 2.1a
{\it 2.1a. Method 1: The age-metallicity-galactocentric distance relation} % section 2.1a
%.........................................................................................Section 2.1a

\bigskip\noindent
The first method to be considered was initially developed by Maciel et al. (\cite{mcu2003}), in the 
framework of an estimate of the time variation of the radial abundance gradients in the galactic disk. 
Using the oxygen abundance measured in the nebula, which is given in the usual form $\epsilon ({\rm O}) = 
\log ({\rm O/H}) + 12$, the [O/H] abundance relative to the Sun is simply given by
$[{\rm O/H}] = \log ({\rm O/H}) - \log({\rm O/H})_\odot$, where we have adopted $\epsilon({\rm O})_\odot = 
8.7$ (see for example Asplund \cite{asplund2003} and Asplund et al. \cite{asplund2004}). The relation 
between the metallicity [Fe/H] and the oxygen abundance is given by

   \begin{equation}
     [{\rm Fe/H}] = 0.0317 + 1.4168 \ [{\rm O/H}] \ ,
     % eq. 1
   \end{equation}

\noindent
which is valid for $-1.0 < [{\rm Fe/H}] < 0.5$, as discussed by Maciel et al. (\cite{mcu2003}). Finally, 
the ages of the PN progenitor stars are given by an age-metallicity-galactocentric distance relation 
developed by Edvardsson  et al. (\cite{edvardsson}),

   \begin{equation}
     \log t = 0.872 - 0.303 \ [{\rm Fe/H}] - 0.038 \ R \ ,
     % eq. 2
   \end{equation}

\noindent
where $t$ is in Gyr, and $R$ is the galactocentric distance in kpc, so that some knowledge of the 
distance to the PN must be assumed. These relationships were applied to a sample of 234 well 
observed PN from Maciel et al. (\cite{mcu2003}), located in the solar neighbourhood and in the 
galactic disk, for which the data are obtained with the highest accuracy. These objects have 
galactocentric distances in the range $4 < R ({\rm kpc}) < 14$, and most (69\%) are located in the 
solar neighbourhood, with distances $d \leq 3\,$kpc.  The results are shown in Fig.~1. It can be 
seen that the distribution shows a prominent peak, located around 3-6 Gyr, similar to the  age of 
the Sun, suggesting that most PN come from stars having masses close to one solar mass when in the 
Main Sequence.

%--------------------------------------------------------------------------
   \begin{figure}
   \centering
   \includegraphics[angle=-90,width=8cm]{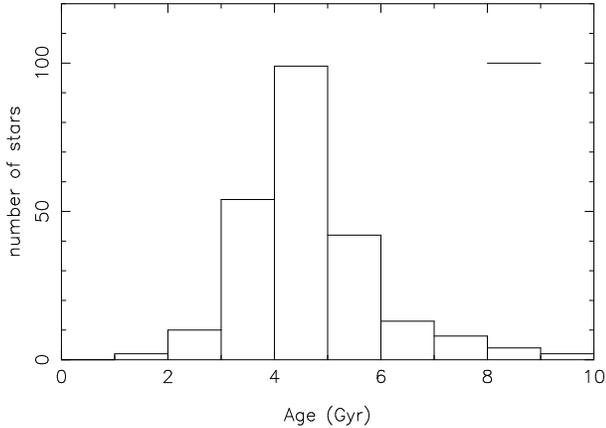}
      \caption{Age distribution of the central stars of planetary nebulae 
      using an age-metallicity-galactocentric distance relation. The estimated
      age uncertainty is shown at the upper right corner.}
      \label{fig1}
      % figura 1
   \end{figure}
%--------------------------------------------------------------------------

\bigskip
%\subsection{Method 3: The age-metallicity relation} % section 2.3

\noindent
%.........................................................................................Section 2.1b
{\it 2.1b. Method 2: The age-metallicity relation for the galactic disk} % section 2.1b
%.........................................................................................Section 2.1b
\bigskip

\noindent
Rocha-Pinto et al. (\cite{rmsf2000}) derived an age-metallicity relation for the galactic disk 
based on chromospheric ages and accurate metallicities. According to this relation (cf. Table~3 of 
Rocha-Pinto et al. \cite{rmsf2000}), we can write

   \begin{equation}
     [{\rm Fe/H}] = 0.13969 - 0.08258 \ t + 0.00277 \ t^2
     % eq. 3
   \end{equation}

\noindent
where $t$ is in Gyr. From this equation the stellar lifetimes can be determined once the metallicity 
is fixed. We can apply the same procedure as in the previous method (cf. Section 2.1a), and obtain 
[Fe/H] from the oxygen abundance $\epsilon({\rm O})$. The results for the same sample considered in 
Fig.~1 are shown in Fig.~2. The derived age distribution is flatter than in the previous method, 
but in both cases most stars have ages under 6~Gyr.

%--------------------------------------------------------------------------
   \begin{figure}
   \centering
   \includegraphics[angle=-90,width=8cm]{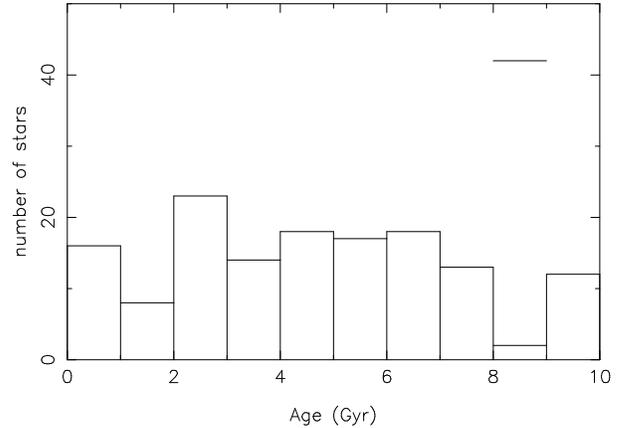}
      \caption{Age distribution of the central stars of planetary nebulae
      using the age-metallicity relation of Rocha-Pinto et al. (\cite{rmsf2000}).
      The estimated age uncertainty is shown at the upper right corner.}
      \label{fig2}
      % figura 2
   \end{figure}
%--------------------------------------------------------------------------

% aqui comeca a discussao dos erros do metodo 2.1a

\noindent
%.........................................................................................Section 2.1c
{\it 2.1c. Uncertainty analysis} % section 2.1c
%.........................................................................................Section 2.1c
\bigskip

\noindent
Oxygen abundances are the best determined abundances in PN, with uncertainties typically under 
0.2 dex, while other elements have generally higher uncertainties, in the 0.2 to 0.3 dex range. 
In the case of the best determined O/H abundances, formal uncertainties under 0.10 dex are often 
estimated. The objects included in our sample are all disk planetary 
nebulae, avoiding the more distant objects for which the abundances are poorly known. They 
result  from a very careful selection, in which the best spectra available were taken into 
account, and for which a comparison of abundances from different sources produces a very good
agreement (see for example the individual abundance comparisons shown in Maciel et al. 
\cite{mlc2006}). All abundances considered are derived from collision excitation lines,
which are considered as true representatives of the ionized gas in the nebulae (cf. Liu 
\cite{liu}). An additional uncertainty may be introduced by the ON conversion which may occur 
in the progenitor stars (cf. Stasi\'nska \cite{stasinska3}). However, present 
results are not conclusive, and in any case they would only affect the progenitor 
stars near the upper mass bracket of the Main Sequence  stars that produce planetary 
nebulae, which are a small fraction of the sample considered here. 

The solar oxygen abundance is accurate within 0.05 dex (see for example Asplund et al.
\cite{asplund3}), so that the uncertainty in the derived [O/H] abundances is essentially
the same as in $\epsilon({\rm O})$. On the other hand, from the correlations presented
in Maciel et al. (\cite{mcu2003}), iron and oxygen are clearly in lockstep in the galactic 
disk, so that we can safely adopt this hypothesis for our present sample. The average
uncertainty in the [Fe/H] metallicity is dominated by the uncertainty in the [O/H]
ratio, which is essentially the same as in the $\epsilon({\rm O})$ abundance, since
the uncertainties in the slope and intercept of Eq.~(1) are small, corresponding to
0.049 and 0.016, respectively. Therefore, an upper limit of about 0.3 dex can be  estimated
for the uncertainty in [Fe/H], but a more realistic average would be about 0.2 dex,
which corresponds to an O/H uncertainty of roughly 0.2 dex.

Concerning the ages as given by Eq.(2), a similar procedure taking into account the 
[Fe/H] uncertainties leads to age uncertainties in the range 0.5 to 1 Gyr for objects
in the solar neighbourhood, which are the majority in the sample considered here.
 
The term involving the galactocentric distance provides a further uncertainty, but the
range estimated above is not significantly changed. The abundances themselves are 
distance-independent, but distance uncertainties may affect the galactic abundance gradients 
and as a secondary effect the position of the nebula projected onto the galactic plane. 
Independent confirmation of the reality of the gradients are presented by Pe\~na et al. 
(\cite{psg}), where a new method to determine  the distances was developed, leading to a clear 
disk gradient that flattens out in the inner galaxy, a result confirmed by Pottasch and
collaborators (cf. Gutenkunst et al. \cite{gutenkunst}). The main uncertainties in the 
gradients were discussed in detail by Maciel et al. (\cite{mlc2005b}), and the effect of 
the adopted distances was also analyzed by Maciel et al. (\cite{mlc2006}) and more recently 
by Maciel \& Costa (\cite{mc2009}). In the latter work, we have taken into account four 
different PN distance scales, our own Basic Sample, and  the scales by Cahn et al. 
(\cite{cks}), Zhang (\cite{zhang}), and Stanghellini et al. (\cite{ssv}). The main 
conclusion is that there is no appreciable change in the gradients when a sample of over 
200 nebulae are considered with central stars in the age range of 2 to 10 Gyr. It was also 
shown that, in average, the galactocentric distances obtained by these scales do not differ 
by more than about 1 kpc for objects in a ring centered at the solar position ($R_0 \simeq 
7.5 - 8.0\,$kpc) and extending to about 3 kpc in either direction, which includes most 
nebulae in our sample. 

In agreement with these results, a typical uncertainty of $\sigma(\log t) \simeq 0.10$, 
where $t$ is in Gyr, was estimated from the analysis by Edvardsson et al. (\cite{edvardsson}), 
on which Eq.~(2) is based. Therefore, the age distribution of Fig.~1 may be displaced by about 
1 Gyr in either direction. Taking into account the observed scatter in the age-metallicity 
relation, this uncertainty is probably larger, up to about 0.2 dex (see a detailed discussion 
in Rocha-Pinto et al. \cite{rmsf2000}, \cite{rp2006}). As a consequence, absolute {\it individual} 
ages may have uncertainties higher than 1 Gyr, but it should be stressed that here we are 
interested in the {\it age distribution}, so that an actual displacement of the histogram of 
Fig.~1 by about 1 Gyr, as indicated by the horizontal bar at the upper right corner of the 
figure, is a realistic estimate.

% aqui comeca a discussao dos erros do metodo 2.1b

Considering now the uncertainties involved in the approximation given by Eq.(3), we would like
to stress that an age-metallicity relation is expected in the framework of a simple model of 
galactic chemical evolution (cf. Prantzos \cite{prantzos}), and in fact several independent 
investigations have been able to derive some working relationships in the last two decades 
(cf. Rocha-Pinto et al. \cite{rmsf2000} and references therein; Bensby et al. \cite {bensby}). 
However, some problems  are still under discussion, particularly the observed dispersion of this 
relation and its applicability to the thick and/or thin disks. The observed dispersion depends critically
on the samples considered, and from the results by Evardsson et al. (\cite{edvardsson}) and 
Feltzing et al. (\cite{feltzing}), a relatively large dispersion of about 0.3 dex was estimated
for the [Fe/H] ratio. Our own results indicate that the actual dispersion may be considerably lower, 
about 0.2 dex (Rocha-Pinto et al. \cite{rmsf2000}, \cite{rp2006}), provided some important corrections 
are made, concerning the cosmic scatter, incompleteness of the samples and a careful analysis of stars 
that present contradictory age indicators, as discussed in detail by Rocha-Pinto et al. (\cite{rcm2002}). 
This value agrees very well with our estimate of the formal uncertainty, as discussed above.
Therefore, the similarity of the uncertainties of the methods discussed in Sections 2.1a and 2.1b 
are reinforced, and the average age uncertainy of 1 Gyr is shown as a horizontal bar
at the upper right corner of Fig.~2. These results are supported by a recent 
analysis by Pranztos (\cite{prantzos}), who presented a detailed discussion on the local 
age-metallicity relation based on the results of the Geneva-Copenhagen Survey (cf. Nordstr\"om 
et  al. \cite{nordstrom}). The resulting relationship shows a decrease in the metallicities 
with increasing ages, as expected, up to about 6 Gyr, remaining essentially flat at later epochs. 
Taking into account the biases affecting this relation by using simulated data as input for the 
age-metallicity relation, a monotonic relation was obtained, up to about 10 Gyr ago, in better 
agreement with the views expressed in our previous work.

\bigskip
%\subsection{Method 2: N/O masses of CSPN} % section 2.2

\noindent
%.........................................................................................Section 2.2
{\it 2.2. Method 3: The age-N/O mass relation for CSPN} % section 2.2
%.........................................................................................Section 2.2
\bigskip

\noindent
%.........................................................................................Section 2.2a
{\it 2.2a. The age-N/O mass relation} % section 2.2a
%.........................................................................................Section 2.2a
\bigskip

\noindent
This method was also employed by Maciel et al. (\cite{mcu2003}), and assumes a relationship between 
the central star mass $m_{CS}$ and the N/O abundance (for details see Maciel et al. \cite{mcu2003}), 
which is expected from several theoretical and observational analyses. The adopted relation can
be written as

   \begin{equation}
     m_{CS} = 0.7242 + 0.1742 \ \log ({\rm N/O}) 
     % eq. 4
   \end{equation}

\noindent
for $-1.2 < \log ({\rm N/O}) < -0.26$, and

   \begin{equation}
     m_{CS} = 0.825 + 0.936  \ \log ({\rm N/O}) +  1.439  \ [\log ({\rm N/O})]^2  
     % eq. 5
   \end{equation}

\noindent
for $-0.26 < \log ({\rm N/O}) < 0.20$. In these equations $m_{CS}$ is in solar masses.
In order to obtain the stellar masses on the main sequence $(m_{MS})$, we  adopted an initial 
mass-final mass relation of the form

   \begin{equation}
     m_{CS} = a + b \ m_{MS} + c \ (m_{MS})^2 \ ,
     % eq. 6
   \end{equation}

\noindent
where $a = 0.47778$,  $b = 0.09028$, and $c = 0$, to reproduce the same values in Maciel et 
al. (\cite{mcu2003}), namely $m_{CS} = 0.55\,M_\odot$ for $m_{MS} = 0.8\,M_\odot$ and 
$m_{CS} = 1.2\,M_\odot$ for $m_{MS} = 8\,M_\odot$, which reflect the known ranges of both
CSPN and their progenitor stars on the main sequence. Concerning the mass-age relation, 
an approximation generally considered as accurate can be written as

   \begin{equation}
     \log t = d + e \ \log m_{MS} + f \ (\log m_{MS})^2 \ .
     % eq. 7
   \end{equation}

\noindent
where $t$ is the Main Sequence lifetime, or age, measured in Gyr, and the Main Sequence mass
$m_{MS}$ is given in solar masses. There are many discussions in the literature which suggest
different values for the constants $d,\,e,\,f$ (see for example Romano et al. \cite{romano}),
so that we will consider two cases here. In the first case, which we call Case A, we have 
$d = 1.0$, $e = -2.0$ and $f = 0.0$, which corresponds to a relation of the form
$t = C/m_{MS}^2$, where $C = 10\, {\rm M_\odot^2} \, {\rm Gyr}$. The second case, which we
call Case B, adopts the well known mass-age relation by Bahcall \& Piran (\cite{bp1983}), 
which  corresponds to  $d = 1.0$, $e = -3.6$, and $f = 1.0$ in Eq.~(7). Again we have 
$t = 10\,$Gyr for $m_{MS} = 1 M_\odot$. Taking into account Cases A and B has two advantages:
first, it stresses that this important relation still involves some uncertainties;
and second, the relation considered as Case B gives a steeper decrease of the stellar
lifetimes at larger masses compared to Case A, so that we can easily interpret the
behaviour of the age distributions as a function of the mass-age relationship.

Method 3 was applied to a  sample of 122 PN for which all necessary data was available 
(cf. Maciel et al. \cite{mcu2003}). Again we selected disk planetary nebulae for
which the best abundance data were available so as to keep the uncertainties to a minimum. 
The results  are shown in Figs. 3, and~4 for cases A and B, respectively. These results are 
similar to the age-metalllicity-radius method, in the sense that most objects have ages lower 
than about 10 Gyr, and there is a sharp maximum in the probability distribution. However, its location 
depends on the estimated lifetimes as a function of the Main Sequence mass, being around
3-7 Gyr for Case A and 1-4 Gyr for Case B. From case~A to case~B the lifetimes of the 
more massive stars are decreased, so that the probability of finding stars at larger lifetimes 
decreases, and the whole peak moves to the left, as shown by Figs.~3 and~4.

%--------------------------------------------------------------------------
   \begin{figure}
   \centering
   \includegraphics[angle=-90,width=8cm]{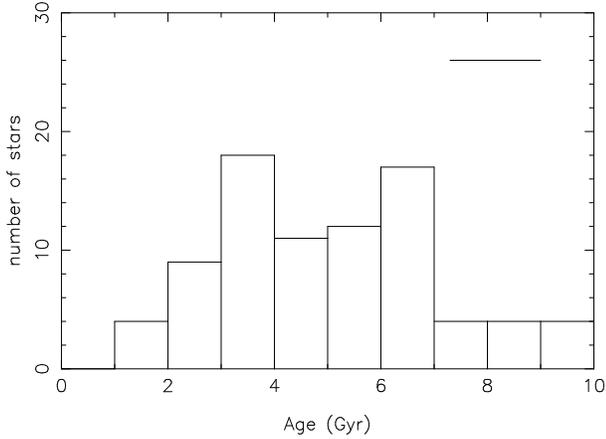}
      \caption{Age distribution of the central stars of planetary nebulae
      using a mass-N/O abundance relation (Case A). The estimated age 
      uncertainty is shown at the upper right corner.} 
      \label{fig3}
      % figura 3
   \end{figure}
%--------------------------------------------------------------------------

%--------------------------------------------------------------------------
   \begin{figure}
   \centering
   \includegraphics[angle=-90,width=8cm]{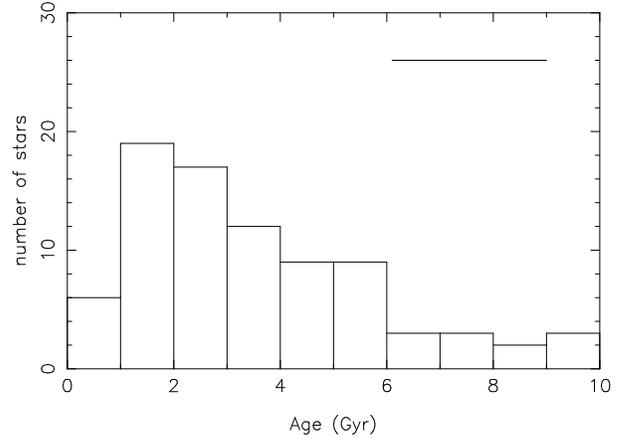}
      \caption{The same as Fig.~3 for Case B.}
      \label{fig4}
      % figura 4
   \end{figure}
%--------------------------------------------------------------------------

\bigskip

\noindent
%.........................................................................................Section 2.2b
{\it 2.2b. Uncertainty analysis} % section 2.2b
%.........................................................................................Section 2.2b
\bigskip

\noindent
An estimate of the uncertainties involved in Method 3 suggests that they are similar or 
somewhat larger than in Methods~1 and 2. The basic relation is given by Eqs.(4) and (5), which 
relate the central star masses and the N/O abundances. Although such a relation is 
expected from theoretical models (see for example Renzini \& Voli \cite{renzini}, Marigo 
\cite{marigo2000}, \cite{marigo2001}, Perinotto et al. \cite{perinotto}), there is  presently no 
{\it clearcut} functional dependence between  these quantities, which led Cazetta \& Maciel 
(\cite{cazetta}) and Maciel (\cite{m2000}) to propose a  calibration based on the best parameters 
available.  The N/O ratio is relatively well determined, with uncertainties similar to the O/H ratio 
or even better, namely, under 0.2 dex, as already discussed. Adopting this value and also taking
into account the uncertainties in the coefficients of Eqs.~(4) and (5), an upper limit of about 
0.2 $M_\odot$ would be obtained for the mass uncertainty. However, a more realistic estimate 
would be lower than this value, as suggested by the recent application of N/O masses to a well 
studied sample of CSPN by Maciel et al. (\cite{mkc2008}). 

In that work, the derived masses were successfully compared with the known mass distributions of CSPN 
and white dwarf stars in order to explain the relationship between the modified momentum of the CSPN 
winds and the stellar luminosity.  Some additional support to the N/O masses comes from the reanalysis 
of Tinkler \& Lamers (\cite{tinkler}) of the central star masses determined by Kudritzki et al. 
(\cite{kudritzki}). These results were based on an homogeneous set of parameters obtained from 
Zanstra temperatures, dynamical ages and evolutionary tracks.  The average masses obtained by 
Tinkler \& Lamers (\cite{tinkler}) for the same sample studied by Maciel et al. (\cite{mkc2008}) 
and the N/O masses are in very good agreement. In view of these considerations, an average 
uncertainty closer to $0.1\,M_\odot$ would be more appropriate to Eqs.~(4) and (5).

Considering now the initial mass-final mass relation as given by Eq.(6), there are many 
determinations of this relation in the literature (Vassiliadis \& Wood \cite{vw1993}, 
Marigo \cite{marigo2000}, \cite{marigo2001}, Marigo \& Girardi \cite{marigo2007}, Meng et al. 
\cite{meng}), which generally agree with each other within 0.1 to 0.2 $M_\odot$, so that the 
total uncertainties in the stellar masses are probably not much affected by this relation. As 
a consequence, the average uncertainties in the {\it indivdual} CSPN ages correspond to a mass 
uncertainty of at most 0.2 $M_\odot$. In fact, the main contribution to the age uncertainty is 
due to the age estimates as given by Eq.(7), and at this point the best procedure to overcome 
such difficulty is to adopt some kind of parametrization, by considering cases~A  and B. Applying 
the mass uncertainty mentioned above, a formal age uncertainty  would be about 3.1 and 4.7 
Gyr for cases A and B, respectively, considering a typical star of one solar mass at the Main 
Sequence. These value can be considered as upper limits, as they were obtained using a $0.2\,M_\odot$
mass uncertainty. Adopting the more reasonable value of about $0.1\,M_\odot$, the average 
uncertainties would be 1.7 Gyr and 2.9 Gyr for Cases  A and B, as shown by the error bars in
Figs.~3 and 4. Again, it should be stressed that we are interested in the {\it age distribution}, 
rather than in individual ages, so that the general effect of the uncertainties is expected to 
be smaller than indicated by the formal uncertainties.

\bigskip

\noindent
%.........................................................................................Section 2.2c
{\it 2.2c. Binary CSPN} % section 2.2c
%.........................................................................................Section 2.2c
\bigskip

\noindent
A final comment is appropriate on Method~3, concerning the possibility that a significant fraction
of the PN may have originated from close binary stars. Current results are controversial,
in the sense that a very small number of binary PN are known, as discussed by De Marco 
(\cite {demarco09}), while some investigations suggest that binaries constitute a much larger 
proportion of the galactic nebulae (see for example Miszalski et al. \cite{brent}, De Marco 
\cite{demarco06}, \cite{demarco09}, Moe \& De Marco \cite{moe}). In view of these contradictory 
aspects, it is difficult to evaluate the effect of binarity on the present results. It would be 
expected that the age distribution would be displaced towards higher ages by only a small amount, 
as suggested by the good agreement of the results of our methods with the actual mass distribution 
of CSPN and white dwarf stars, as we will see in the next section.

%--------commentary on the SMC planetary nebulae -----------------------------------------------

%In an investigation of the chemical evolution of the Small Magellanic Cloud  (SMC) based on 
%abundance data of planetary nebulae, Idiart et al. (\cite{imc2007}) obtained ages for a sample 
%of SMC PN, using high quality data to derive the properties of the progenitor stars. Theoretical 
%evolutionary tracks from Vassiliadis \& Wood (\cite{vw1992}) have been used, and  a large number 
%of measured spectral fluxes for each nebula has been collected, so that  accurate physical 
%parameters could be derived. The obtained age distribution is similar to the one of Method~2, case B,
%in the sense that most objects have ages under 4 Gyr, and the peak occurs for $t \sim 1-2\,$Gyr.
%However, the SMC sample is very small and a comparison of SMC objects with data from our 
%own Galaxy cannot be directly made, as the SMC is younger than the Galaxy and has a larger
%star formation rate, so that one would expect the average ages of the CSPN to be lower than in 
%the Galaxy. This is indeed observed, so that we may consider these ages as lower limits to the 
%ages of the galactic CSPN.

%-----------------------------------------------------------------------------------------------------
\section{The expected age distribution of the central stars of planetary nebulae} % section 3
%-----------------------------------------------------------------------------------------------------

\subsection{The white dwarf mass distribution} % section 3.1

The expected age distribution of the CSPN can be estimated by analyzing the much better known mass 
distribution of the white dwarf stars. Since the average mass loss rates during the PN phase amount 
to about $dM/dt \sim 10^{-8}$ to $10^{-6}\, M_\odot/{\rm yr}$ and the whole PN phase duration is 
about $\Delta t \sim 1$ to $2  \times 10^4\, {\rm yr}$, the total mass lost during this phase is 
$\Delta m \sim (dM/dt)\, \Delta t \sim 10^{-4}$ to $2 \times 10^{-2}\, M_\odot$,
which is much smaller than the CSPN masses. Therefore, as a first approximation, the mass distribution 
of the CSPN must be similar to that of the white dwarfs, except for the very low mass stars with 
$m < 0.55\,M_\odot$. Such stars are not expected from theoretical models, since main sequence stars 
leading to white dwarfs with masses lower than about $0.55\,M_\odot$ probably go directly to the white 
dwarf phase.

Recent work on the mass distribution of white dwarfs by Madej et al. (\cite{madej}) and Kepler et al. 
(\cite{kepler}) lead to a distribution which is strongly peaked at about $0.56\,M_\odot$, as shown 
in Fig.~5 (cf. Madej et al. \cite{madej}). This investigation was based on a large sample of about 1200 
white dwarfs, and shows very few objects (about 5\%) with masses larger than about $0.8\,M_\odot$.

%--------------------------------------------------------------------------
   \begin{figure}
   \centering
   \includegraphics[angle=-90,width=8cm]{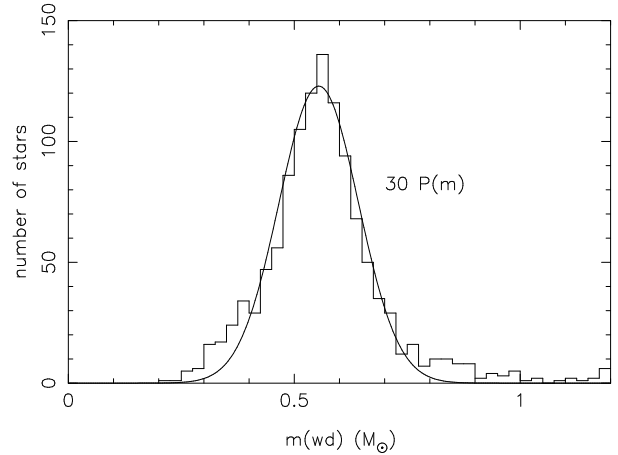}
      \caption{Mass distribution of white dwarfs (Madej et al. \cite{madej})
       and Gaussian fit.}
      \label{fig5}
      % figura 5
   \end{figure}
%--------------------------------------------------------------------------

The white dwarf mass distribution can be well fitted by a Gaussian probability density distribution 
defined by

   \begin{equation}
     P(m) = {N(m) \over N_t} \ {1 \over \Delta m}
     % eq. 8
   \end{equation}

\noindent
where $N(m)$ is the number of stars with mass $m$, $N_t$ is the total number of stars, and 
$\Delta m = 0.025\,M_\odot$ is the size of the adopted mass  bins, as shown in Fig.~5, where 
the curve plotted is $30 \times P(m)$. As can be seen, the gaussian fit to the data is 
good, although at very small and very large masses the predictions are somewhat lower
than observed. Such probability is normalized, that is, 
$\int P(m) \ dm = 1$. For white dwarfs $P(m)$ can then be written as

   \begin{equation}
   P(m) =  {A \over \sigma \ \sqrt{2\pi}} \ e^{-{1\over 2}\bigl({m - \mu\over \sigma}\bigr)^2} \ ,
     % eq. 9
   \end{equation}

\noindent
where $A = 0.90349$, $\sigma = 0.08798$ and $\mu = 0.55420$. Assuming that the CSPN have 
approximately  the same mass distribution than the white dwarfs, we have
$P_{CS}(m)\ dm = P_{WD}(m) \ dm$.
As a first approximation, we can assume a linear relation between the CSPN mass and the Main Sequence (MS) 
mass, as given by Eq.~(6) with $c = 0$, so that we have then $a = 0.47778$ and $b = 0.09028$,
as before. The main sequence masses can be written as

   \begin{equation}
   m_{MS} = {1 \over b} \ (m_{CS} - a) 
     % eq. 10
   \end{equation}

\noindent
and $dm_{CS} = b \ dm_{MS}$, so that the probability distribution for MS stars can be written as
$P_{MS}(m)\ = b \ P_{CS}(m)$.
The derived probability distribution of MS stars is shown in Fig.~6. It can be seen that it peaks 
around one solar mass, and also includes some very low (even negative) masses, which of course are 
unrealistic, as they correspond to CSPN masses lower than $m_{CS} = a \simeq 0.48\,M_\odot$, which are 
not observed. However, as we will see later on, this essentially increases the probability of very 
high lifetimes, an effect that can be interpreted with some modifications in the calculation of the 
probability $P_{MS}(m)$.

%--------------------------------------------------------------------------
   \begin{figure}
   \centering
   \includegraphics[angle=-90,width=8cm]{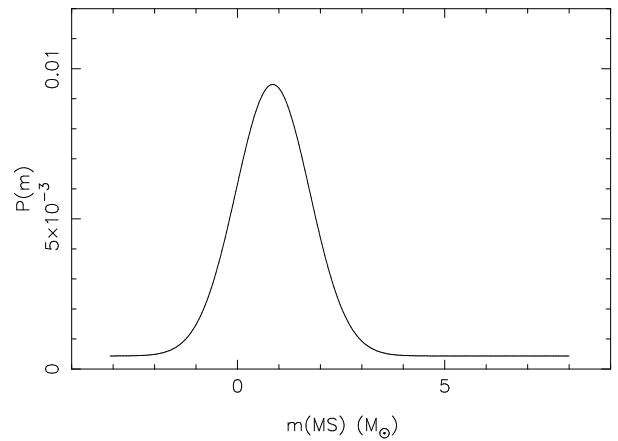}
      \caption{Mass distribution of CSPN progenitor stars at the Main Sequence.}
      \label{fig6}
      % figura 6
   \end{figure}
%--------------------------------------------------------------------------

Assuming that the star formation rate in the Galaxy has remained approximately constant along the 
galactic lifetime, the age distribution of the CSPN can be estimated from the mass distribution of 
their progenitor stars. In order to derive the age distribution of the observed CSPN, we will adopt 
cases A and B considered previously. Considering that $P(t)\ dt = P_{MS}(m)\ dm$, we have for case A:

   \begin{equation}
   P(t) = {m^3 \over 2C} \ P_{MS}(m)  \ ,
     % eq. 11
   \end{equation}

\noindent
and  for case B:

   \begin{equation}
   P(t) = {m^{4.6} \over \vert 2\,\log m - 3.6 \vert} \ {1 \over 10^{1+(\log m)^2}} \ P_{MS}(m)
     % eq. 12
   \end{equation}

\noindent
where we have dropped the subscript MS from the mass. The obtained age distributions are shown in 
Fig.~7 for Case A (solid curve) and B (dashed curve), respectively. Case A gives  larger lifetimes 
for masses greater than one solar mass, which pushes the peak of the probability distribution to 
the right (2-3 Gyr), while for case B these lifetimes are shorter, and the peak moves to the left
($\sim 1$ Gyr). Excluding the MS stars that do not lead to the formation of CSPN, the main effect 
on Fig.~7 is a sharp decrease in the probability for ages greater than about 12 Gyr, leaving the 
peak region essentially unaffected. The main conclusion that can be drawn is that a peaked 
distribution can be expected, but the precise location of the peak depends on the adopted 
initial mass-final mass relation and especially the calculation of the lifetimes. 

%--------------------------------------------------------------------------
   \begin{figure}
   \centering
   \includegraphics[angle=-90,width=8cm]{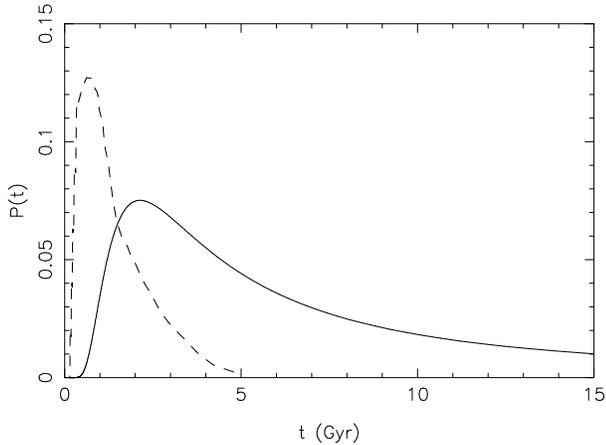}
      \caption{Age distribution of the progenitors of the central stars of planetary 
      nebulae. Solid line: case A, dashed line: Case B.}
      \label{fig7}
      % figura 7
   \end{figure}
%--------------------------------------------------------------------------

\noindent
For the sake of generality, we have also considered a quadratic %m_{CS} = f(m_{MS})$
relation as in Eq.~(6) with $c \neq 0$ instead of a linear equation. In this case, we have adopted 
the constants $a$,  $b$, and $c$ from the recent work of Meng et al. (\cite{meng}), taking the 
metal abundance $Z = 0.02$. The central star mass is given by the minimum of $m_{CS}(a_1,b_1,c_1)$ 
and $m_{CS}(a_2,b_2,c_2)$, where $a_1 = 0.5716$, $b_1 = -0.04633$, $c_1 = 0.02878$, $a_2 = 1.1533$, 
$b_2 = -0.2422$, and $c_2 = 0.04091$. In this case, the probability $P_{MS}(m)$ is given by
$P_{MS}(m) = (b + 2 \, c\ m_{MS}) \ P_{CS}(m)$. As it turned out, the results of Fig.~7 are 
not particularly sensitive to this assumption.

\subsection{The observed mass distribution of the central stars of planetary nebulae} % section 3.2

The mass distributions of CSPN and white dwarfs have also been previously considered by Stasi\'nska 
et al. (\cite{stasinska}) and Napiwotzki (\cite{napiwotzki}). More recently, Gesicki \& Zijlstra 
(\cite{gesicki}) analyzed these distributions based on a dynamical method which allows mass 
determinations within $0.02\,M_\odot$ (cf. Gesicki et al. \cite{gesicki1}). The CSPN masses were
obtained for a sample of 101 objects from a relation between the temperatures of the central stars 
and the dynamical age of the surrounding nebulae. The ages were derived from a combination of recent 
spectra, line ratios and nebular sizes, using a photoionization model to obtain the central star 
temperature. Theoretical tracks by Bl\"ocker (\cite{blocker}) were used to derive the stellar masses. 
It results that both the CSPN and white dwarf distributions peak around $0.6\,M_\odot$ as in Madej 
et al. (\cite{madej}) and Stasi\'nska et al. (\cite{stasinska}), although the white dwarf distribution 
shows a broader mass range. The CSPN distribution shows essentially no objects with masses 
higher than $0.7\,M_\odot$, while in the case of the DA white dwarfs, which are presumably the 
offspring of H-rich CSPN, a few objects are observed with  masses larger than $0.8\,M_\odot$. 
Although there may be some differences between the recently obtained white dwarf mass distributions, 
as discussed by Gesicki \& Zijlstra (\cite{gesicki}), they all agree in the sense that any sizable 
sample of  CSPN is expected to have a larger number of objects with masses close to $0.6\,M_\odot$. 
These results are in good agreement with our own N/O masses, as discussed by Maciel et al. 
(\cite{mkc2008}).

Instead of adopting the white dwarf mass distribution, we may then use directly the mass 
distribution of  CSPN as recently derived by Gesicki and Zijlstra (\cite{gesicki}). The mass 
distribution is shown in Fig.~8, and can also be approximated as a Gaussian probability density 
distribution similar to eq.~(9), where we find $A = 0.89627$, $\sigma = 0.01431$, and 
$\mu = 0.6087$. The corresponding probability distribution for CSPN is also shown in Fig.~8. 
Adopting  again the same hypotheses as in Section~3.1, the CSPN age distributions are shown  
in Fig.~9 for Case A (solid curve) and B (dashed curve), respectively. In this case, the peaks are
located at approximately 4-6 Gyr (Case A) and 2-4 Gyr (Case B).

%-----------------------------------------------------------------------------------------------------
\section{Discussion} % section 4
%-----------------------------------------------------------------------------------------------------

Some indication of the age distribution of CSPN may be obtained from the PN classification 
originally proposed by Peimbert (\cite{peimbert1}, \cite{peimbert2}). According to this 
classification, the following approximate ages are attributed to the different PN types: 
Type I (1 Gyr), Type II (3 Gyr), Type III (6 Gyr), and Type IV (10 Gyr) (see for example 
Stasi\'nska \cite{stasinska2}, table 7). Type IV are halo objects, which are rarer and 
presumably much older than the remaining types, which are located in the disk and bulge 
of the Galaxy. Since nebulae of Types I-III constitute the vast majority of the known galactic 
PN, in this work we will not take into account the older, Type IV nebulae.

%--------------------------------------------------------------------------
   \begin{figure}
   \centering
   \includegraphics[angle=-90,width=8cm]{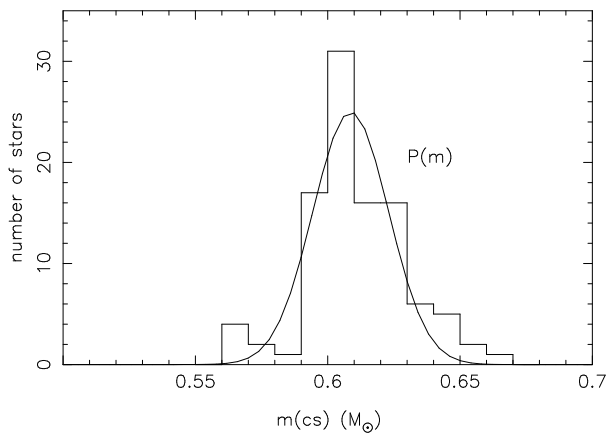}
      \caption{Mass distribution of the central stars of planetary nebulae 
      from Gesicki \& Zijlstra (\cite{gesicki})
      and Gaussian fit.}
      \label{fig8}
      % figura 8
   \end{figure}
%--------------------------------------------------------------------------

Fig.~7 shows the CSPN age distributions for Cases A and  B using the mass distribution of white 
dwarfs (Madej et al. \cite{madej}), and Fig.~9 gives the corresponding distributions from the CSPN 
(Gesicki and Zijlstra \cite{gesicki}) mass distribution. The merged age distributions suggest 
a maximum around 2-6 Gyr for Case A and 1-4 Gyr for case B. Again, the main conclusion that can be 
drawn is that a peaked distribution can be expected, but the location of the peak depends on the 
adopted assumptions. In principle, we would expect Case B to be more realistic than Case A (see 
the discussion on stellar lifetimes by Romano et al. \cite{romano}), so that the results of Section~3 
would suggest a preferable range of 1-4 Gyr for the peak of the distribution. On the other hand, 
the results based on the empirical CSPN mass distribution are probably more accurate than those 
inferred from the white dwarf mass distribution, for the reasons mentioned in Section 3. Therefore, 
in view of these considerations, and taking into account the uncertainty analyses of Sect.~2.1c.
and 2.2b., we suggest that the peak of the age distribution is probably located around 2-4 Gyr, 
as shown by the dashed curve of Fig.~9. We  may then compare this distribution with the results 
of Sect.~2. Taking into account the uncertainties involved in the age determinations, which are
estimated to be in the range 1.0-2.0 Gyr, approximately, as discussed in Sect.~2, it can be 
concluded that all methods considered produce results reasonably in agreement with the expected age 
distribution. According to the discussion in Sect.~2, Methods 1 and 2 have similar results, as
indicated by Figs.~1 and 2, but the age distribution of Method 2 is less strongly peaked than
either of the distributions obtained in Sect.~3 (cf. Figs.~7 and 9). Since the uncertainties are
similar, Method~1 is probably more accurate than Method~2.

At face value, the age distributions of Method~3 shown in Figs.~3 and 4 are very similar to the
results of Fig.~9, especially if Case B is considered, as seems more appropriate. In other words,
the best results are probably those by Method~3, Case B (Fig.~4), which show a very good agreement
with the results from the empirical mass distribution of CSPN, as shown by the dashed curve of
Fig.~9. Although the estimated uncertainty of Method~3 is larger, it is reassuring that the
average distribution shows such a remarkable agreement with the expected age distribution of CSPN.
Therefore, our results obtained from completeley independent methods and sources of data are 
reasonably in agreement with each other, so that we can have an estimate of the age distribution 
of CSPN. Naturally, the details of the individual age determinations still need to be worked out.

%--------------------------------------------------------------------------
   \begin{figure}
   \centering
   \includegraphics[angle=-90,width=8cm]{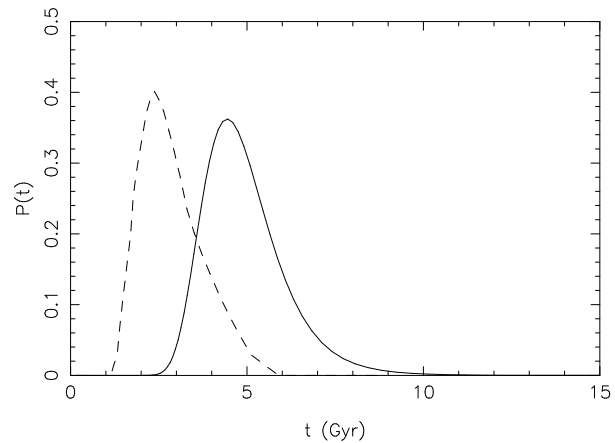}
      \caption{Age distribution of the CSPN progenitor stars from the observed 
       mass distribution. Solid line: Case~A, dashed line: Case~B.}
      \label{fig9}
      % figura 9
   \end{figure}
%--------------------------------------------------------------------------

\begin{acknowledgements}
      We are indebted to an anonymous referee, whose detailed comments helped
      improving the present paper. This work was partially supported by FAPESP and CNPq. 
\end{acknowledgements}


\begin{thebibliography}{}


   \bibitem[2003]{asplund2003} Asplund, M. 2003, CNO in the universe, ed. C. Charbonnel, 
   D. Schaerer, G. Meynet, ASP CS 304, 275

   \bibitem[2004]{asplund2004} Asplund, M., Grevesse, N., Sauval, A. J., Allende-Prieto, C., 
   \&  Kiselman, D. 2004, A\&A, 417, 751

   \bibitem[2006]{asplund3} Asplund, M., Grevesse, N., Sauval, A. J. 2006, Nuclear Phys. A, 777, 1

   \bibitem[1983]{bp1983} Bahcall, J. N., \& Piran, T.  1983, ApJ,  267, L77

   \bibitem[2004]{bensby} Bensby, T., Feltzing, S., \& Lundstr\"om, I. 2004, A\&A, 421, 969

   \bibitem[1995]{blocker} Bl\"ocker, T. 1995, A\&A, 299, 755

   \bibitem[1992]{cks} Cahn, J. H., Kaler, J. B., \& Stanghellini, L. 1992, A\&AS, 94, 399 

   \bibitem[2000]{cazetta} Cazetta, J. O., \& Maciel, W. J. 2000, Rev. Mex. A\&A, 36, 3 

   \bibitem[2006]{demarco06} De Marco, O. 2006, IAU Symp. 234, ed. M. J. Barlow, \& R. H. M\'endez,
        CUP, 111

   \bibitem[2009]{demarco09} De Marco, O. 2009, PASP, 121, 316

   \bibitem[1993]{edvardsson} Edvardsson, B., Anderson, J., Gustafsson, B., et al. 1993, 
   A\&A, 275, 101

   \bibitem[2001]{feltzing} Feltzing, S., Holmberg, J., \& Hurley, J. R. 2001, A\&A, 377, 911

   \bibitem[2007]{gesicki}  Gesicki, K., \& Zijlstra, A. A. 2007, A\&A, 467, L29 

   \bibitem[2006]{gesicki1}  Gesicki, K., Zijlstra, A. A., Acker, A., et al. 2006, A\&A, 451, 925 

   \bibitem[2008]{gutenkunst} Gutenkunst, S., Bernard-Salas, J., Pottasch, S. R.,
        Sloan, G. C., \& Houck, J. R.  2008, ApJ, 680, 120

   \bibitem[2009]{hillebrand} Hillebrand, L., Mamajek, E., Stauffer, J., Soderblom,
       D., Carpenter, J., \& Meyer, M. 2009, Cool stars, stellar systems and the Sun,
       15th. Cambridge Workshop, AIP Conference Proceedings Vol. 1094, 800

   \bibitem[2007]{imc2007} Idiart, T. P., Maciel, W. J., \& Costa, R. D. D. 2007, 
    A\&A,  472, 101

   \bibitem[2007]{kepler} Kepler, S. O. et al., 2007, MNRAS,  375, 1315

   \bibitem[1997]{kudritzki} Kudritzki, R.-P., M\'endez, R.~H., Puls, J.,  \& McCarthy, 
     J.~K. \ 1997, IAU Symp.180, Eds. H.~J. Habing, \& H.~J.~G.~L.~M. Lamers, Kluwer, 64

   \bibitem[2006]{liu} Liu, X.-W. 2006, IAU Symp. 234, ed. M. J. Barlow, \& R. H. M\'endez,
        CUP, 219

   \bibitem[2000]{m2000} Maciel, W. J. 2000, IAU Symposium 198, ed. L. da Silva,
        M. Spite, J. R. de Medeiros, ASP, 204

   \bibitem[2009]{mc2009} Maciel, W. J.,  \& Costa, R. D. D. 2009, The Milky Way and
       the Local Group: Now and in the GAIA Era, Heidelberg Conference

   \bibitem[2003]{mcu2003} Maciel, W. J.,  Costa, R. D. D., \& Uchida, M. M. M. 
   2003, A\&A,  397, 667

   \bibitem[2008]{mkc2008} Maciel, W. J.,  Keller, G. R., \& Costa, R. D. D. 2008, 
   Rev. Mex. A\&A, 44, 221

   \bibitem[2005a]{mlc2005a} Maciel, W. J., Lago, L. G., \& Costa, R. D. D. 2005a, 
   A\&A, 433, 127

   \bibitem[2005b]{mlc2005b} Maciel, W. J., Lago, L. G., \& Costa, R. D. D. 2005b, 
      in Planetary Nebulae as Astronomical Tools, ed. R. Szczerba, G. Stasi\'nska,
      \& S. K. Gorny, AIP, CP 804, 246

   \bibitem[2006]{mlc2006} Maciel, W. J., Lago, L. G., \& Costa, R. D. D. 2006, 
   A\&A, 453, 587

   \bibitem[2004]{madej} Madej, J., Nalezyty, M., \& Althaus, L. G. 2004, 
   A\&A, 419, L5

   \bibitem[2000]{marigo2000} Marigo, P. 2000, in The evolution of the Milky Way,
      ed. F. Matteucci, F. Giovanelli, Kluwer, 481

   \bibitem[2001]{marigo2001} Marigo, P. 2001, A\&A, 370, 194

   \bibitem[2007]{marigo2007} Marigo, P., \& Girardi, L. 2007, A\&A, 469, 239
    
   \bibitem[2008]{meng} Meng, X., Chen, X., \& Han, Z.  2008, A\&A,  487, 625

   \bibitem[2009]{brent} Miszalski, B., Acker, A., Moffat, A. F. J., Parker, Q. A., \&
        Udalski, A. 2009, A\&A, 496, 813

   \bibitem[2006]{moe} Moe, M., \& De Marco, O. 2006, ApJ, 650, 916

   \bibitem[2006]{napiwotzki} Napiwotzki, R. 2006, A\&A, 451, L27

   \bibitem[2004]{nordstrom} Nordstr\"om, B., Mayor, M., Andersen, J., et al. 2004, A\&A, 418, 989

   \bibitem[1976]{peimbert1} Peimbert, M. 1976, IAU Symp. 76, 
    ed. Y. Terzian, Reidel, Dordrecht, 215

   \bibitem[1990]{peimbert2} Peimbert, M. 1990, Rep. Prog. Phys., 53, 1559

   \bibitem[2004]{perinotto} Perinotto, M., Morbidelli, L., \& Scatarzi, A. 2004, MNRAS, 349, 793

   \bibitem[2005]{psg} Pe\~na, M., Stasi\'nska, G., \& Gorny, S. K. 2005, 
      in Planetary Nebulae as Astronomical Tools, ed. R. Szczerba, G. Stasi\'nska,
      \& S. K. Gorny, AIP, CP 804, 243

   \bibitem[2008]{prantzos} Prantzos, N. 2008, IAU Symp. 254, ed. J. Andersen, J. Bland-Hawthorn,
     \& B. Nordstr\"om, CUP, 381

   \bibitem[1981]{renzini} Renzini, A., \& Voli, M. 1981, A\&A, 94, 193

   \bibitem[2002]{rcm2002} Rocha-Pinto, H. J., Castilho, B. V., \& Maciel, W. J. 
       2002, A\&A, 384, 912

   \bibitem[2000]{rmsf2000} Rocha-Pinto, H. J., Maciel, W. J., Scalo, J., \& Flynn, C. 
   2000, A\&A, 358, 850

   \bibitem[2006]{rp2006} Rocha-Pinto, H. J., Rangel, R. H. O., Porto de Mello, G. F., 
      Bragan\c ca, G. A., \& Maciel, W. J. 2006, A\&A, 453, L9

   \bibitem[2005]{romano} Romano, D., Chiappini, C., Matteucci, F., \& Tosi, M. 
    2005, A\&A, 430, 491

    \bibitem[2009]{soderblom} Soderblom, D. R. 2009, IAU Symp. 258, ed. E. Mamajek, 
      D. R. Soderblom, \& R. Wyse, Cambridge University Press, Cambridge (in press)

   \bibitem[2008]{ssv} Stanghellini, L., Shaw, R. A., \& Villaver, E. 2008, ApJ, 689,  194

   \bibitem[2004]{stasinska2} Stasi\'nska, G. 2004, in Cosmochemistry: The melting pot 
       of the elements, ed. C. Esteban, R. J. Garc\'\i a L\'opez, A. Herrero, \& 
       F. S\'anchez (Cambridge University Press, Cambridge) 115

   \bibitem[2008]{stasinska3} Stasi\'nska, G. 2008, in Stellar Nuclecosynthesis:
       50 years after  B2FH, ed. C. Charbonnel, J.-P. Zahn, EAS Publ. Ser. 32, 173

   \bibitem[1997]{stasinska} Stasi\'nska, G., Gorny, S. K., \& Tylenda, R. 1997, 
   A\&A, 327, 736

   \bibitem[2002]{tinkler} Tinkler, C. M., \& Lamers, H. J. G. L. M. 2002, A\&A, 384, 987

   \bibitem[1993]{vw1993} Vassiliadis, E., \& Wood, P. R. 1993, ApJ, 413, 641 

   \bibitem[1995]{zhang} Zhang, C. Y. 1995, ApJS, 98, 659 



\end{thebibliography}
\end{document}